\documentclass[12pt]{article}
\usepackage{graphicx,tabularx,epsfig}
\usepackage{color}
\usepackage{enumitem}
\usepackage{caption}
\usepackage{subcaption}
\usepackage{amsmath}
\usepackage{amssymb}
\usepackage{amsthm}
\usepackage{physics}
\usepackage{dsfont, mathtools}
\usepackage{psvectorian}
\usepackage{blkarray}
\usepackage{bm}
\usepackage{url}
\usepackage{setspace}

\usepackage{xcolor}
\colorlet{linkequation}{blue}

\usepackage[colorlinks = true,
            linkcolor = blue,
            urlcolor  = blue,
            citecolor = blue,
            anchorcolor = blue]{hyperref}

\usepackage{floatrow}

\newlength{\abstractwidth}
\renewcommand{\thefootnote}{\fnsymbol{footnote}}
\renewcommand{\thanks}[1]{\footnote{#1}} 
\newcommand{\starttext}{
\setcounter{footnote}{0}
\renewcommand{\thefootnote}{\arabic{footnote}}}

\makeatletter
\g@addto@macro\normalsize{%
  \setlength\abovedisplayskip{10pt}
  \setlength\belowdisplayskip{20pt}
  \setlength\abovedisplayshortskip{10pt}
  \setlength\belowdisplayshortskip{20pt}
}
\makeatother

\usepackage{color}

\renewcommand{\title}[1]{\vbox{\center\LARGE{#1}}\vspace{5mm}}
\renewcommand{\author}[1]{\vbox{\center#1}\vspace{5mm}}
\newcommand{\address}[1]{\vbox{\center\em#1}}
\newcommand{\email}[1]{\vbox{\center\tt#1}\vspace{5mm}}

\begin{document}

\singlespacing

\begin{titlepage}
\rightline{}
\bigskip
\bigskip\bigskip\bigskip\bigskip
\bigskip

\begin{center}

{\Large \bf {{Petz map and Python's lunch}}
}
\end{center}

\bigskip \noindent

\bigskip

\begin{center}

\author{Ying Zhao}

\address{Institute for Advanced Study,  Princeton, NJ 08540, USA }

\email{zhaoying@ias.edu}

\bigskip

\vspace{1cm}

\end{center}

\begin{abstract}

We look at the interior operator reconstruction from the point of view of Petz map and study its complexity. We show that Petz maps can be written as precursors under the condition of perfect recovery. When we have the entire boundary system its complexity is related to the volume / action of the wormhole from the bulk operator to the boundary. When we only have access to part of the system, Python's lunch appears and its restricted complexity depends exponentially on the size of the subsystem one loses access to.

\medskip
\noindent
\end{abstract}

\end{titlepage}

\starttext \baselineskip=17.63pt \setcounter{footnote}{0}

{\hypersetup{hidelinks}
\tableofcontents
}

\section{Introduction}

In AdS/CFT correspondence, the bulk operators can be written as boundary operators \cite{Hamilton:2006az}. For an operator inside the causal patch of the boundary region, one can use HKLL and write the operator as a smeared operator on the boundary. For a state with a dual geometry which contains a black hole, it was argued that outside the black hole horizon, the complexity of operator reconstruction is related to the radial location of the bulk operator \cite{Susskind:2014rva}. As an exterior operator approaches the black hole horizon, the complexity blows up. On the other hand, the operator reconstruction beyond causal wedge and inside entanglement wedge is still possible \cite{Jafferis:2015del}\cite{Dong:2016eik}\cite{Faulkner:2017vdd}. In this paper, we look at the interior operator reconstruction from the point of view of Petz map and study its complexity. 

\begin{figure}[H] 
 \begin{center}                      
      \includegraphics[width=2in]{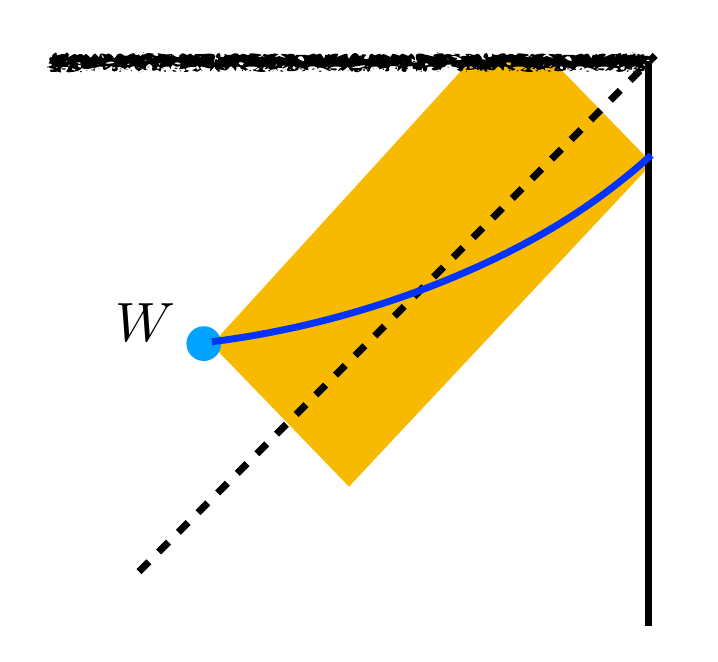}
      \caption{ }
    \label{operator0}
  \end{center}
\end{figure}

Petz map assumes fine-tuned knowledge about the state, i.e., one needs to know how the code subspace is embedded into the the boundary Hilbert space \cite{Almheiri:2014lwa}\cite{Cotler:2017erl}\cite{Chen:2019gbt}. It was argued that the spacetime region in the black hole interior corresponds to the quantum circuit preparing the state \cite{Hartman:2013qma}\cite{Susskind:2014moa}. In the rest of the paper we will call it the horizontal circuit as it grows in spacelike direction. To write down an interior operator with Petz map, one needs to know the portion of the circuit lying between the operator and the black hole horizon (Figure \ref{operator0}). Once one knows the horizontal circuit, one can shorten the wormhole, apply the operator, then evolve back. We show that this is essentially what Petz map does in case of perfect recovery.

Petz map also allows one to reconstruct the operator from part of the boundary system so the operator reconstruction is not unique. The complexity of the reconstructed operator is the lowest when we keep the entire system. In this case, the complexity is given by the volume / action of the part of the interior from the bulk operator to the boundary (Figure \ref{operator0}). In this sense, the relation between radial location and operator complexity as advocated in \cite{Susskind:2014rva} still holds. The deeper a bulk operator is, the more complex its boundary reconstruction is . If one wants to reconstruct the operator from part of the boundary system, the reconstruction is still possible but the complexity can be significantly higher. This is exactly the Python's lunch situation as discussed in \cite{Brown:2019rox}. See also \cite{Harlow:2013tf}. The tensor network complexity is still controlled by the wormhole volume but the circuit complexity is exponential in the number of qubits one loses access to.

Naively the above discussion contradicts the statement that the complexity of a bulk operator blows up as we move it closer and closer to the horizon from the exterior \cite{Susskind:2014rva}. The resolution is that HKLL reconstruction as discussed in \cite{Susskind:2014rva} does not make use of fine-tuned information about the state as Petz map does. If one wants to reconstruct the interior operator using Petz map, one needs fine-tuned information about the horizontal circuit. In general it is very complex to figure this out from the boundary state \cite{Bouland:2019pvu}. That's why the interior reconstruction is hard and this is the essential difference between operator reconstruction inside and outside the horizon. In other words, the horizontal circuit stored in the interior may not be very complex but it is very difficult to figure out what this circuit is from the boundary data.

The rest of the paper is organized as follows. In section \ref{precursors} we look at some examples. Assume perfect recovery we write Petz map in the form of a precursor $UWU^{\dagger}$ for some unitary operator $U$. In section \ref{complexity} we look at its complexity. We compare it with the action / volume from the operator to the boundary when we have access to the entire system. We see the appearance of Python's lunch when we only have part of the boundary system. 
We compare with the operator reconstruction outside horizon using HKLL.

\section{Petz maps as precursors}
\label{precursors}

In this section we study Petz map in various examples under the condition of perfect recovery. We will work in black hole models where the black hole is represented by $S$ qubits. Our goal is to write the Petz map reconstruction of black hole interior operators in the form $UWU^{\dagger}$ for $U$ unitary and $W$ simple operator. In this paper we will use the word ``precursor" in a flexible way and call such an operator a precursor. $U$ may or may not be the black hole time evolution. In this form one can easily analyze its complexity and make a connection with the Einstein-Rosen bridge geometry. 

In this paper we will use the following definition of Petz map from \cite{Chen:2019gbt}\cite{Penington:2019kki}. $W$ is an operator on the code subspace. $V$ is the embedding of the code subspace into the CFT Hilbert space.  Assume we have access to subsystem A and lose subsystem $\bar A$. The reconstruction of $W$ on subsystem $A$ is given by
\begin{align}
\label{Petz_map}
	W_A = \sigma_A^{-\frac{1}{2}}\tr_{\bar A}(VWV^{\dagger})\sigma_A^{-\frac{1}{2}}
\end{align}
where $\sigma_A = \tr_{\bar A}\Pi_{code} = \tr_{\bar A}VV^{\dagger}$.

\subsection{perturbed thermofield double}
Consider the simplest example where Alice perturbs thermofield double by throwing in some object with energy much higher than Hawking temperature. We model this as the following. We represent thermofield double as $S$ Bell pairs shared between Alice and Bob. Then we throw in an extra spin in state $\ket{a}$ to Alice side. We have initial state
\begin{align*}
	\ket{\psi_0} = \frac{1}{\sqrt{N}}\sum_I\ket{a, I}_A\ket{I}_B
\end{align*}
Alice time evolves it for a while, the state becomes
\begin{align*}
	\ket{\psi_1} = \frac{1}{\sqrt{N}}\sum_{I, i}\ket{ i}_AU_{i, aI}\ket{I}_B
\end{align*}
There is a two-dimensional code subspace spanned by the two spin states $\ket{a}$.
\begin{align*}
	V:\ &\mathcal{H}_{code}\longrightarrow \mathcal{H}_A\times\mathcal{H}_B\\
	&\ket{a}\mapsto \frac{1}{\sqrt{N}}\sum_{I,  i}\ket{ i}_AU_{i, aI}\ket{I}_B
\end{align*} 

At this point Alice wants to flip the spin, i.e., map $\ket{a}\rightarrow \sum_b\ket{b}W_{ba}$. A straightforward evaluation of Petz map shows that Alice needs to apply  $UWU^{\dagger}$. We leave the details to Appendix \ref{Petz}. An immediate question is, what is $U$? We don't know $U$. All we know are the following matrices when we trace out Bob's system in the code subspace.\footnote{In this paper, this is what we mean when we talk about knowing the horizontal circuit. We need to know how some particular code subspace we are interested in is embedded the boundary Hilbert space. This is a time-dependent concept. As the wormhole gets longer the embedding becomes more complex. }
\begin{align}
\label{matrices}
	\tr_B(V\ket{a}\bra{b}V^{\dagger}) = \frac{1}{N}\sum_{i, j}\ket{ i}\bra{j}\sum_I U_{i, aI }U^*_{j, bI}
\end{align}

These matrices \eqref{matrices} do not uniquely determine $U$, but they are enough to tell us what is $UWU^{\dagger}$. In this example, we get a precursor $UWU^{\dagger}$ where $U$ is the black hole time evolution. Intuitively, as time increases the black hole interior gets longer while the spin stays inside. Alice undoes the evolution stored in the horizontal circuit by $U^{\dagger}$, gets the spin out, flips it, and then applies $U$ again.

\subsection{Pure state black hole}

Next, we consider pure state black holes. The initial state is $\ket{a, I}$ where $a$ represents a spin. Under time evolution, this state becomes
\begin{align*}
	\psi_1 = \sum_{i,\alpha}\ket{i,\alpha}U_{i\alpha, aI}
\end{align*}
where $\alpha$ represents the subsystem we will trace out later. 

Our task is to flip the spin $a$, i.e., apply $W = \sigma_x$ on the spin. We have one logical qubit embedded into the boundary Hilbert space.
\begin{equation}
\label{embedding1}
\begin{aligned}
	V:&\mathcal{H}_{code}\rightarrow \mathcal{H}\\
	  &\ket{a}\mapsto \sum_{i,\alpha}\ket{i,\alpha}U_{i\alpha, aI}\equiv \ket{e_{aI}}
\end{aligned}
\end{equation}


 We assume we have the knowledge of the following matrices.
\begin{align*}
	\tr_{\bar A}\qty(V\ket{a}\bra{b}V^{\dagger})	=\ &\sum_{\alpha}\ket{i}\bra{j}U_{i\alpha, aI}U^*_{j\alpha , aI}
\end{align*}

\begin{itemize}

\item{Keep the entire system}

First, we consider the simplest case when we don't trace out anything, i.e., $\bar A$ (subsystem corresponding to index $\alpha$) is empty set. We can guess the answer to be $UWU^{\dagger}$. We check this from the definition of the Petz map. In this case, the embedding map $V$ is given in $\eqref{embedding1}$.
\begin{align*}
	V = \sum_i U_{i, aI}\ket{i}\bra{a}
\end{align*}
In the definition of Petz map,
\begin{align*}
	VWV^{\dagger} =\ & \sum_{i,j} \ket{i}U_{i, aI}W_{ab} U^*_{j, bI}\bra{j}\\
	=\ &\sum_{a,b}\ket{e_{aI}}W_{ab}\bra{e_{bI}}\ \ \ \ \ \text{no summation on $I$}
\end{align*} 
Notice that this is NOT the simple matrix product $ UWU^{\dagger}$ as there is no summation on the index $I$. $VWV^{\dagger}$ only acts non-trivially on the the code subspace. Outside the code subspace, it gives zero.

On the other hand, we look at the matrix product $ UWU^{\dagger}=\sum_{a,b,J}\ket{e_{aJ}}W_{ab}\bra{e_{bJ}}$. Comparing with $VWV^{\dagger}$ it has extra summation on the index $J$. 
Notice that we have
\begin{align*}
	&\Pi_{code} = VV^{\dagger} = \sum_{a}\ket{e_{aI}}\bra{e_{aI}}\\
	&VWV^{\dagger} = \Pi_{code}^{\frac{1}{2}}\qty(UWU^{\dagger})\Pi_{code}^{\frac{1}{2}} 
\end{align*} 
As a result, one can write the Petz map as
\begin{align*}
	W_A = \Pi_{code}^{-\frac{1}{2}}(VWV^{\dagger})\Pi_{code}^{-\frac{1}{2}} = UWU^{\dagger}
\end{align*}
Notice that from the knowledge of the code subspace, we know $U_{i, aI}$ for fixed $I$ and any $a$, but this is not enough to determine $UWU^{\dagger}$. In this case, the Petz map \eqref{Petz_map} is not uniquely defined outside the code subspace. Outside the code subspace, $VWV^{\dagger}$ gives zero and $\Pi_{code}$ also gives zero. Their ratio is not uniquely determined. In this example, we simply used the original black hole time evolution to extend it.\footnote{Generally in the definition of Petz map $\sigma_A^{-\frac{1}{2}}$ is defined to annihilate the kernel of $\sigma_A$ so there is no ambiguity. In order to get an unitary operator from Petz map we didn't adopt this convention here. I thank the referee for pointing this out. } 

\item{Tracing out part of the system}

If we don't trace out any part, we can simply reverse the time evolution. If we trace out part of the system, one can still reverse the effect of $U$ but in a more complex way. Here, we assume $U$ is random enough that we can treat it as a perfect tensor, i.e., there is an isometry from a smaller subset of indices to a larger subset. We will show that we could still write the Petz map as $\tilde U W\tilde U^{\dagger}$ for some other unitary operator $\tilde U$. 

We have the state 
\begin{align*}
	\psi_1 = \sum_{i,\alpha}\ket{i,\alpha}U_{i\alpha, aI}
\end{align*}
Subsystem A corresponds to the index $i$ and subsystem $\bar A$ corresponds to index $\alpha$. Let $d_{\bar A}$ be the dimension of subsystem $\bar A$. Now we only have access to subsystem A and we want to flip the spin as before. 

Assuming $|i|>|a||\alpha|$, with fixed index $I$ the tensor $U$ is an isometry from index $a\alpha$ to index $i$. We first extend this isometry to an unitary. We divide the index $I$ into two parts: $I = I_1I_0$ such that $|i| = |a||\alpha||I_1|$. 

 We let
\begin{align*}
	\tilde U_{i, a\alpha I_1'}\equiv d_{\bar A} U_{i\alpha, a I_1' I_0}
\end{align*}

Note that the definition of $\tilde U$ depends on $I_0$. By assumption $\tilde U$ is an unitary operator. Here is a pictorial representation of the operator $\tilde U$. The arrows represent the directions of $U$ and $\tilde U$. 

\begin{figure}[H] 
 \begin{center}                      
      \includegraphics[width=3in]{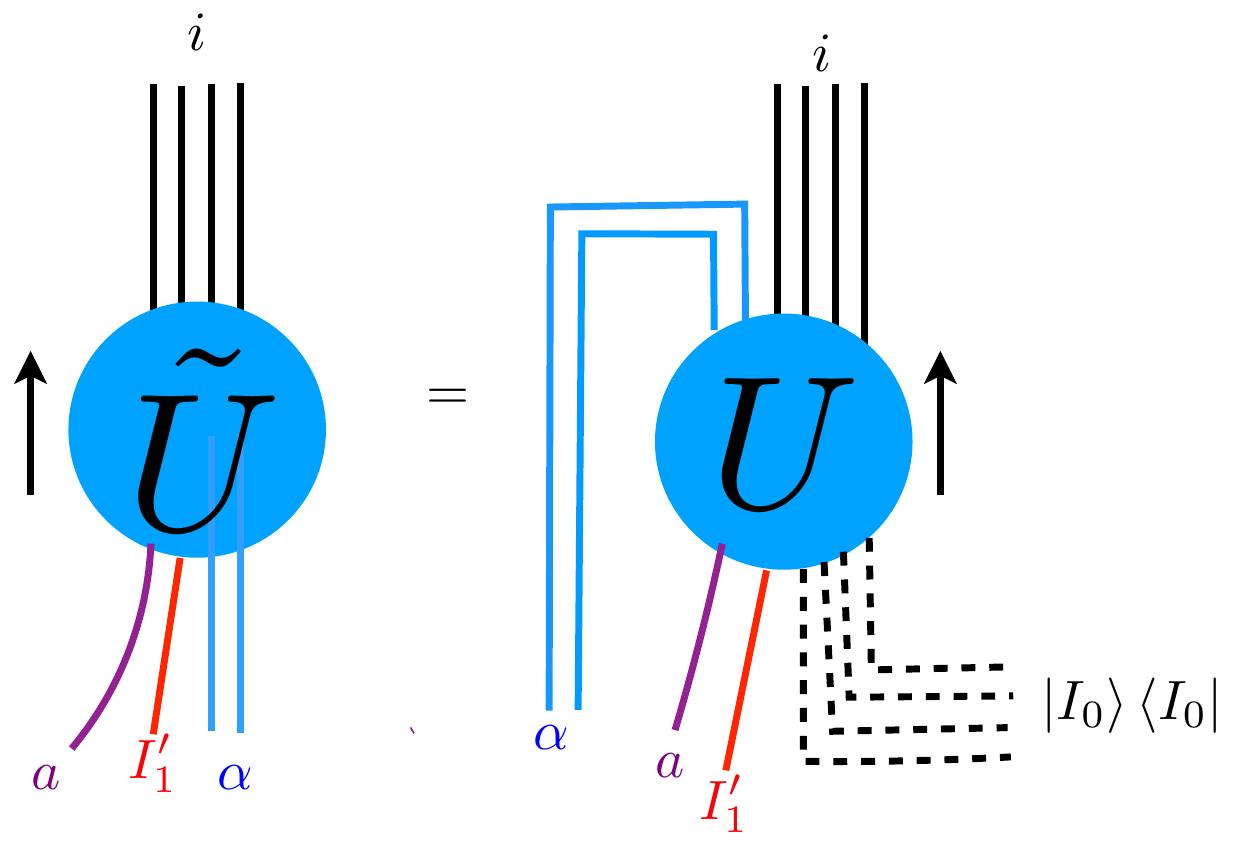}
      \caption{Operator $\tilde U$ involves side-way evolution of operator $U$.  }
    \label{operator}
  \end{center}
\end{figure}

Following earlier notation we let 
\begin{align*}
	\ket{e_{a\alpha I_1' }}\equiv\sum_i \ket{i}\tilde U_{i,a\alpha I_1'}
\end{align*}
By unitarity of $\tilde U$, $\{\ket{e_{a\alpha I_1' }}\}$ from an orthonormal basis in the Hilbert space of subsystem A.

We will show that Petz map can be written as $\tilde U W\tilde U^{\dagger}$.

In the definition of Petz map,
\begin{align*}
	\tr_{\bar A}\qty(VWV^{\dagger}) =\ & \sum_{i,j}\sum_{a,b}\sum_{\alpha}\ket{i}_A\bra{j}U_{i\alpha, aI_1I_0}W_{ab}U^*_{j\alpha , bI_1I_0}\\
	=\ &\frac{1}{d_{\bar A}}\sum_{i,j}\sum_{a,b}\sum_{\alpha}\ket{i}_A\bra{j}\tilde U_{i,a\alpha I_1}W_{ab}\tilde U^*_{j,b\alpha I_1}\\
	=\ &\frac{1}{d_{\bar A}}\sum_{a,b,\alpha}\ket{e_{a\alpha I_1}}W_{ab}\bra{e_{b\alpha I_1}}\ \ \ \ \text{no summation over $I_1$}
\end{align*}
Again, notice that we have summation over $\alpha$ but we don't have summation over $I_1$. This is not a precursor. 

The projection part is given by
\begin{align*}
	&\sigma_A=\tr_{\bar A}(VV^{\dagger})\\
	 =\ & \sum_{a,\alpha}\sum_{i,j} \ket{i}_A\bra{j}U_{i\alpha, a I_1 I_0}U^*_{j\alpha, a I_1 I_0}=\frac{1}{d_{\bar A}}\sum_{a,\alpha}\ket{e_{a\alpha I_1}}\bra{e_{a\alpha I_1}}\ \ \ \ \text{no summation over $I_1$}
\end{align*}

On the other hand, the precursor $\tilde UW\tilde U^{\dagger}$ is given by
\begin{align*}
	\tilde U W \tilde U^{\dagger}
	=\ &\sum_{i,j}\sum_{a,b}\sum_{\alpha, I_1'}\ket{i}_A\bra{j}\tilde U_{i,a\alpha I_1'}W_{ab}\tilde U^*_{j, b\alpha I_1'}\\
	=\ &\sum_{a,b,\alpha,I_1'}\ket{e_{a\alpha I_1'}}\mathcal{O}_{ab}\bra{e_{b\alpha I_1'}}\ \ \text{with summation on $I_1'$}
\end{align*}
We see that
\begin{align}
\label{projection}
	\sigma_A^{\frac{1}{2}}(\tilde UW\tilde U^{\dagger})\sigma_A^{\frac{1}{2}} = \tr_{\bar A}\qty(VWV^{\dagger})
\end{align}
In \eqref{projection}, the projection with $\sigma_A$ gets rid of the summation on $I_1'$. So we can write the Petz map as
\begin{align*}
	W_A = \sigma^{-\frac{1}{2}}\tr_{\bar A}(VWV^{\dagger})\sigma_A^{-\frac{1}{2}}=\tilde UW\tilde U^{\dagger}
\end{align*}

Here is the circuit representation $W_A\ket{\psi_1}$.

\begin{figure}[H] 
 \begin{center}                      
      \includegraphics[width=6in]{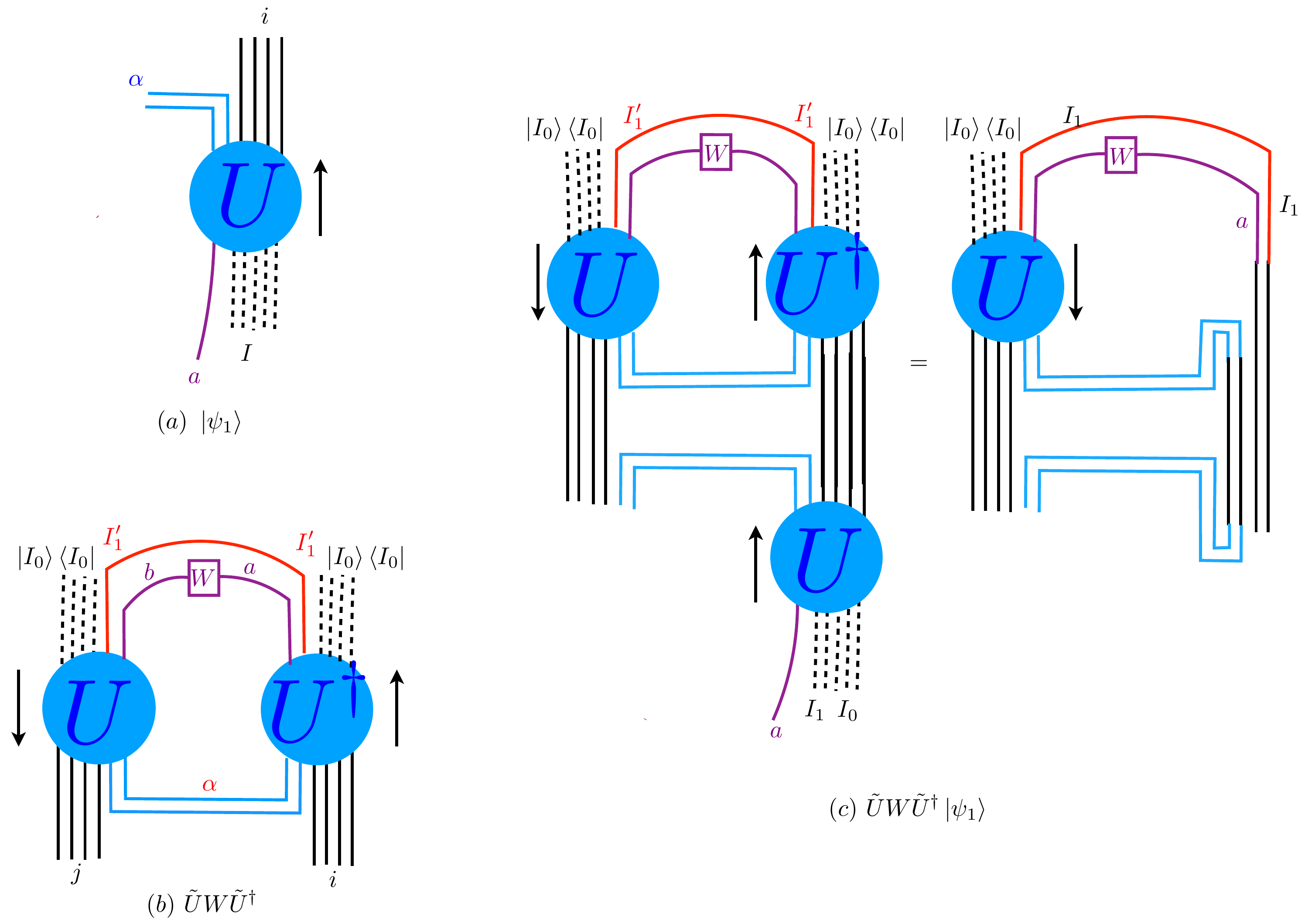}
      \caption{(a) represents the state $\ket{\psi_1}$. (b) represents the operator $W_A = \tilde UW\tilde U^{\dagger}$. (c) represents $W_A\ket{\psi_1}$. }
    \label{circuit2}
  \end{center}
\end{figure}


\end{itemize}

\subsubsection{Comments on the applicability and state dependence}

In this section we discuss the following question: What is the class of pure black hole microscopic states for which such operator reconstruction is applicable?\footnote{I thank the referee for asking these questions.} What's crucial in such reconstruction is that we know how the code subspace is embedded into the CFT Hilbert space through the embedding map $V:\mathcal{H}_{code}\rightarrow \mathcal{H}_{CFT}$. The interior operator is constructed such that it works in the code subspace. 

In our discussion, we only applied Petz map to the simplest case where the embedding from the bulk code subspace to the boundary Hilbert space is essentially given by black hole time evolution. One can easily generalize the discussion to the case where the time evolution is perturbed by various shockwaves. For more general states like Haar-typical black hole microstates with a given mass, we don't know what the interior geometry looks like and we certainly don't know if there exists an embedding of the bulk code space into the larger boundary Hilbert space. The discussion in this paper doesn't apply. 

There have been extended discussions about state dependence of the interior operator reconstruction \cite{Almheiri:2013hfa}\cite{Papadodimas:2013jku}. In our discussion, as the reconstructed operator has the form $\tilde UW\tilde U^{\dagger}$, it is manifestly state-independent within the entire code subspace. This is because we assumed exact error correction and also focused on the case where the dimension of the code subspace is much smaller than the dimension of the black hole Hilbert space. The state dependence in more general case is outside the scope of this paper and see section 3 of \cite{Penington:2019npb} for extended discussion on this issue.

\section{Complexity of the Petz map and Python's lunch}
\label{complexity}

We wrote Petz map in the form of $\tilde U W\tilde U^{\dagger}$. In this section we look at its complexity. 

\subsection{Complexity of Petz map and wormhole geometry}

In earlier examples, when we keep the entire boundary system, the Petz maps simply gives $UWU^{\dagger}$ where $U$ is the Hamiltonian time evolution and is $k$-local. The complexity of such operators was studied in \cite{Stanford:2014jda}\cite{Susskind:2014jwa}\cite{Brown:2016wib}. It grows exponentially before scrambling time and linearly after scrambling time. A simple epidemic model gives its time dependence as 
\begin{align}
\label{epidemic}
	\mathcal{C}(UWU^{\dagger}) = S \log(1+\frac{\delta S}{S}e^{\tau})
\end{align}
where $\tau$ is the circuit time. It is related to the boundary time $t$ as $\tau = \frac{2\pi}{\beta}t$.

As we discussed earlier, the operator $W_A$ essentially undoes the horizontal circuit between the spin and the boundary, applies the operator, then applies the horizontal circuit again. From this point of view, one expects its complexity to match the volume / action from the bulk operator to the boundary. We check this in the following $AdS_2$ geometry. 

\begin{figure}[H] 
 \begin{center}                      
      \includegraphics[width=2in]{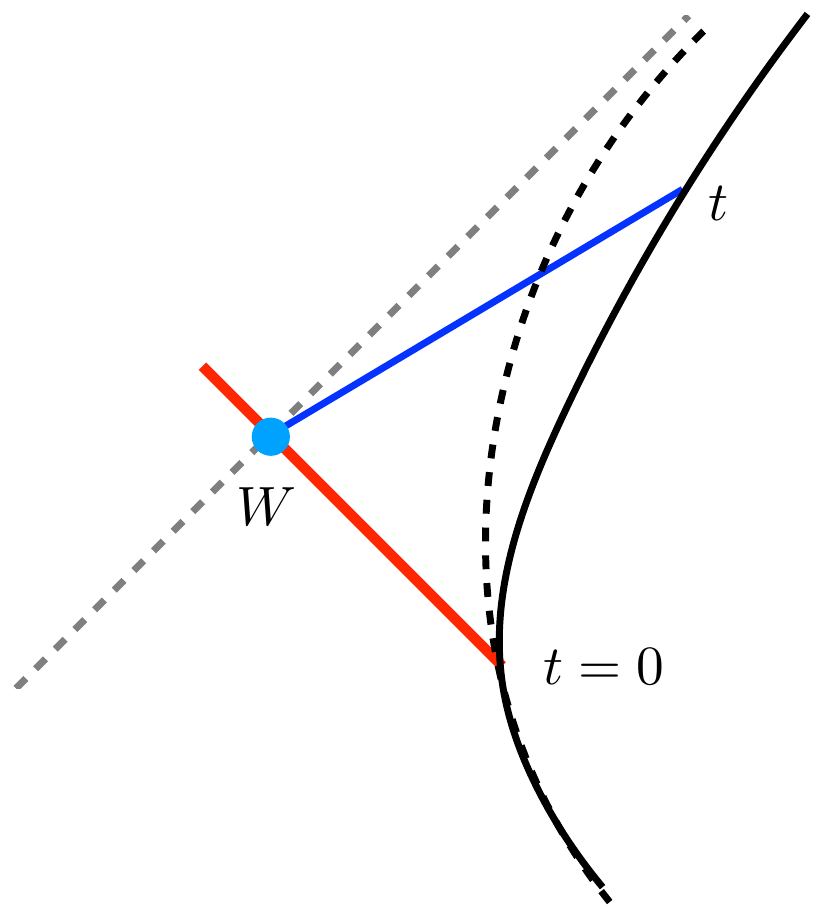}
      \caption{ }
    \label{wormhole_expand}
  \end{center}
\end{figure}

In Figure \ref{wormhole_expand}, the red line represents the spin qubit with index $a$ in earlier discussions. The dashed lines represent the boundary trajectory and the original horizon if the spin qubit were not there. The blue dot is the intersection of the spin world line with the original horizon. The blue line is a geodesic connecting that intersection and the boundary at time $t$. We look at its length. 
In appendix \ref{growth}, we evaluate the geodesic distance in JT gravity and it is given by
\begin{align}
\label{geodesic}
	d(t) \approx \log(1+\frac{\delta S}{S-S_0}\qty[\cosh(\frac{2\pi}{\beta}t)-1])+\text{const}
\end{align}

For $t>\beta$, the time dependence in \eqref{epidemic} matches that in \eqref{geodesic}. One may worry that this is special about $AdS_2$. In \cite{Zhao:2017isy} a spacetime volume evaluation was done in BTZ black hole and one gets the same result. 

We make some comment about the above calculations. We want to look at the geodesic distance between the operator and the boundary at time $t$. The Wheeler-DeWitt patch at time $t$ includes the world line of the spin. Here, we considered the operator as being located at the intersection of the spin world line with the original horizon. Why is this the reasonable choice? First, the blue line in Figure \ref{wormhole_expand} is inside the portion of the geometry that is affected by the spin. Also, we cannot go deeper than the original horizon for the following reason. In our setup, we specify the initial black hole state by $\ket{a}\ket{I}$. We didn't specify anything about $\ket{I}$ other than being a black hole with $S-1$ qubits. $\ket{I}$ may or may not have a smooth classical interior. For example, it can be the superposition of two black holes with completely different wormhole length. All we know is that $\ket{I}$ has an exterior geometry corresponding to a black hole with entropy $S-1$. As a result we look at the portion of the geometry outside the original horizon after the spin comes in. This region stores the gates that involves the extra spin\cite{Zhao:2017isy}.

\subsection{Python's lunch}
With Petz map one can reconstruct the operator even if we don't have the entire boundary system. However, the complexity will be significantly higher. In this case, the Python's lunch scenario as discussed in \cite{Brown:2019rox} will appear. Consider $W_A=\tilde U W\tilde U^{\dagger}$ where $\tilde U$ is given in Figure \ref{operator}. We assume the black hole contains $S$ qubits and we lose access to $n$ qubits, i.e., $U$ is an unitary operator on $S$ qubits while $\tilde U$ is an unitary operator acting on $S-n$ qubits. $U$ is a $k$-local with relatively low complexity. $\tilde U$ is an operator from one subset of indices of $U$ to another subset. This is essentially Python's lunch.

\begin{figure}[H]
\begin{center}                      
      \includegraphics[width=4in]{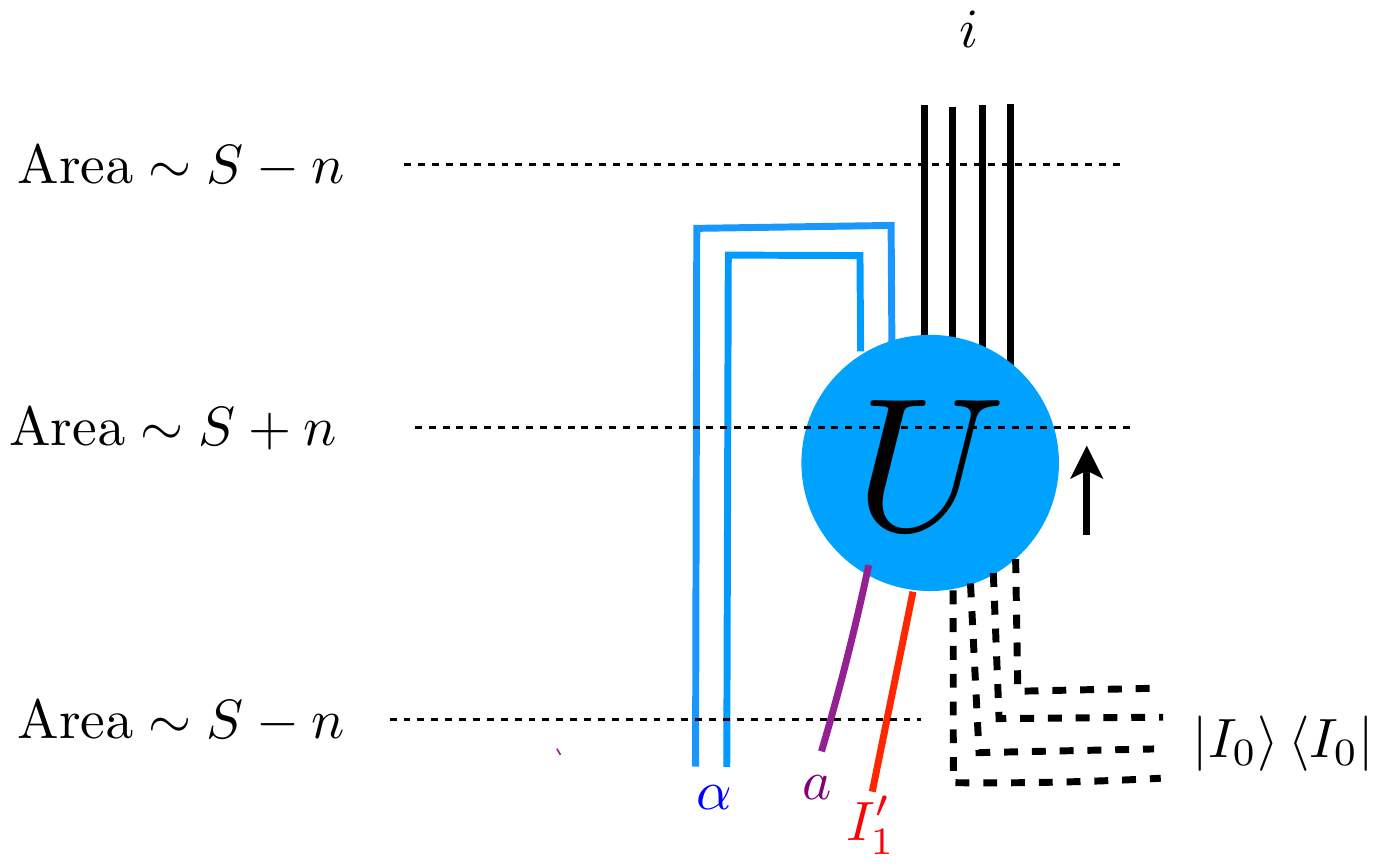}
      \caption{ }
    \label{operator2}
  \end{center}
\end{figure}

To see this, we look at the tensor network representation of operator $\tilde U$. Figure \ref{operator2} illustrates the cross-sectional area of the tensor network at various instances. We see that the minimal cross-sectional area corresponds to $S-n$ qubits while the maximal cross sectional area 
 corresponds to $S+n$ qubits.

In \cite{Brown:2019rox}, it was conjectured that the complexity of unitary circuit $\tilde U$ is controlled by $e^{\frac{1}{2}(S_{max}-S_{min})}$. In this case, it implies complexity $\sim e^n$.

Let's look at the circuit in more detail. We first allow post selection. Here is one way to implement $\tilde U^{\dagger}$. 
Our goal is to realize the following map:
\begin{align*}
	\ket{i}\rightarrow \sum_{b, \beta, I_1'}\ket{b,\beta, I_1'}U^*_{i\beta, bI_1' I_0}\ \ \text{for fixed $I_0$}
\end{align*}
\begin{figure}[H] 
 \begin{center}                      
      \includegraphics[width=2in]{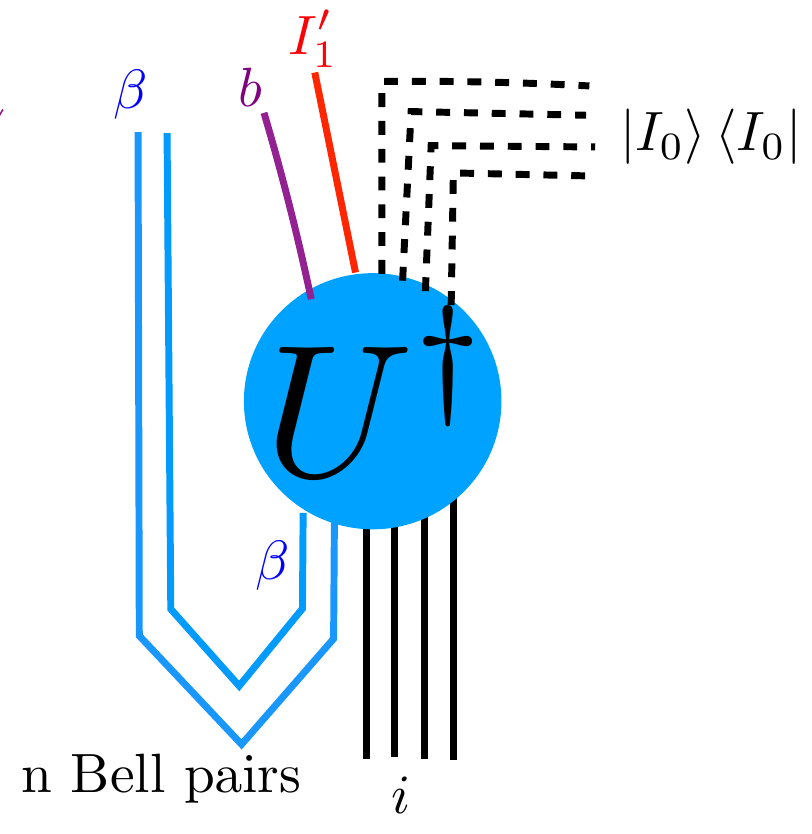}
      \caption{ }
    \label{operator3}
  \end{center}
\end{figure}

In Figure \ref{operator3}, we start from $S-n$ qubits in initial state $i$. We also prepare $n$ extra Bell pairs. We have state 
\begin{align}
\label{initial}
	\ket{\psi_{initial}} = \sum_{\beta}\ket{\beta}\underbrace{\ket{\beta}\ket{i}}_{\text{S qubits}}
\end{align}

 Combining one half the Bell pairs and the initial $S-n$ qubits, together we have $S$ qubits (Equation \eqref{initial}). We apply $U^{\dagger}$ to these $S$ qubits. We get state

\begin{align*}
	\ket{\psi_1} = \sum_{\beta}\ket{\beta}\sum_{b, I_1',I_0'}\ \underbrace{\ket{b,I_1'}}_{\text{$S-2n$ qubits}}\ \ \ \underbrace{\ket{I_0'}}_{\text{$2n$ qubits}}U^*_{i\beta, bI_1' I_0'}
\end{align*}
Now we project the $2n$ qubits corresponding to index $I_0'$ to state $I_0$. We get the state we want.
\begin{align*}
	\qty(\ket{I_0}\bra{I_0})\ket{\psi_1} = \sum_{\beta,b,I_1'}\ \underbrace{\ket{\beta}\ket{b,I_1'}}_{\text{$S-n$ qubits}}U^*_{i\beta, bI_1'I_0}\ket{I_0}
\end{align*}

If we allow post-selection, its tensor network complexity mostly comes from $U$ and is still controlled by the volume / action of the bulk region from the operator to the boundary.

Here we make some comments on this post selection. If we keep the entire system, we have initial state $\sum_{i,\alpha}\ket{i}\ket{\alpha}U_{i\alpha, aI_1I_0}$. Those qubits represented by $i$ (subsystem $A$) and those qubits represented by $\alpha$ (subsystem $\bar A$) are entangled in a particular way such that when evolving backward by $U^{\dagger}$, the extra spin qubit $\ket{a}$ comes out. Now we replace the subsystem $\bar A$ by one half of $n$ Bell pairs that's not entangled with subsystem $A$, i.e., we have initial state
$\sum_{\beta,i,\alpha}\ket{\beta}\qty(\ket{\beta}\ket{i})U_{i\alpha, aI}\ket{\alpha}$. We look at the bulk dual of the subsystem represented by the second $\beta$ and subsystem A represented by $i$. As the two subsystems are unentangled, their dual will be a disconnected pair of spacetime but there is certain probability that they are connected. We evolve backward in time by $U^{\dagger}$. When we do post selection, we project onto the part of the wavefunction for which the two regions are initially connected. As the probability of success in projection is $e^{-2n}$, one expects that initially there is probability $e^{-2n}$ for the two regions to be connected.

In \cite{Yoshida:2017non}\cite{Brown:2019rox} it was shown that instead of doing post selection, one can apply unitary operator to achieve this, but one needs to run the circuit $U$ $\mathcal{O}(e^{n})$ times. It gives unitary circuit complexity $\mathcal{C}(U)e^n$ as predicted by Python's lunch.

In our discussion we assumed $\sigma_A$ has flat spectrum. \cite{Gilyen:2020gmg} gave an algorithm to implement Petz map without such assumptions. Their algorithm is based on Grover search and essentially gives the same result on complexity.\footnote{I thank the referee for bringing this paper to my attention.}

\subsection{Comparison with exterior operator reconstruction}

One may be surprised that it's not so complex to apply an operator in the interior. Here, we need to differentiate between two concepts, the complexity to figure out an operator and the complexity to apply an operator once it is known. Petz map is based on the knowledge of the code subspace, i.e., the knowledge of the horizontal circuit lying between the bulk operator and the boundary. In \cite{Bouland:2019pvu} it was shown that it can be very complex to figure out the horizontal circuit. Here is one of their setups. Consider a subset of boundary states that can be generated by perturbing the time evolution of thermofield double up to time $T$ with $l\ll S$ simple operators, i.e., the horizontal circuits are black hole time evolution perturbed at various points and the perturbations are separated by more than scrambling time. Their claim is that it would be very difficult (complexity at least $e^S$) to distinguish such states from Haar-random states. 

We've seen that in order to apply Petz map to reconstruct the interior operator, one needs knowledge of this horizontal circuit. So it can be very complex to figure out the operator reconstruction. The complexity of doing this can be at least exponential in the complexity of the operator\cite{Bouland:2019pvu}\cite{Susskind:2020kti}. However, once the horizontal circuit is given, the reconstructed operator itself may not be that complex depending on how many qubits one loses access to. 

 On the other hand, to reconstruct an exterior operator from HKLL, no such fine-tuned knowledge is needed. For example, with a black hole background, all one needs to know is the mass, charge e.t.c  of the black hole. This is the essential difference between the interior and exterior operator reconstruction.

 \section*{Acknowledgments}

I thank Ahmed Almheiri, Adam Bouland, Daniel Harlow, Adam Levine, Juan Maldacena, Geoff Penington, Douglas Stanford
 for helpful discussions.
 I thank KITP for hospitality when this work was done. I am supported by the Simons foundation through the It from Qubit Collaboration.

\appendix

\section{Petz map of perturbed thermofield double}
\label{Petz}

Start from the state
\begin{align*}
	\psi_0 = \frac{1}{\sqrt N}\sum_I\ket{a,I}_A\ket{I}_B
\end{align*}
After the time evolution, it becomes
\begin{align*}
	\psi_1 = \frac{1}{\sqrt N}\sum_I\sum_i\ket{i}_AU_{i, aI}\ket{I}_B
\end{align*}
The encoding is given by
\begin{align*}
	V:\ &\mathcal{H}_{code}\rightarrow \mathcal{H}_{CFT}\\
	&\ket{a}\rightarrow \frac{1}{\sqrt N}\sum_I\sum_i\ket{i}_AU_{i, aI}\ket{I}_B
\end{align*}

\begin{align*}
	\tr_{B}(VWV^{\dagger}) = \ &\frac{1}{N}\tr_B\qty(\ket{i, I}\bra{i', I'}U_{i, aI}W_{ab}U^*_{i', bI'})\\
	=\ &\frac{1}{N}\sum_{i,j}\ket{i}\bra{j}\sum_{I, a, b}U_{i, aI}W_{ab}U^*_{j, bI}
\end{align*}

\begin{align*}
	\sigma_A = \tr_B(VV^{\dagger}) = \frac{1}{N}\sum_{i,j}\ket{i}\bra{j}\sum_{I, a}U_{i, aI}U^*_{j, aI} = \frac{1}{N}\mathds{1}_A
\end{align*}
The Petz map gives
\begin{align*}
	W_A = \sigma_A^{-\frac{1}{2}}\tr_B(VWV^{\dagger})\sigma_A^{-\frac{1}{2}} = UWU^{\dagger}
\end{align*}

\section{Growth of the wormhole}
\label{growth}
We compute the growth of the wormhole after a perturbation in JT gravity. We write $AdS_2$ as a hyperboloid in $\mathds{R}^{2,1}$. In embedding coordinates, it is given by $-(Y^{-1})^2-(Y^0)^2+(Y^1)^2 = -1$ . The metric in $R^{2,1}$ is given by $ds^2 = -(dY^{-1})^2-(dY^0)^2+(dY^1)^2$.  The boundary particle trajectory satisfies given by $X^2 = 0$, $\dot X^2 = -1$. \cite{Maldacena:2017axo}

Without perturbation, the right boundary trajectory is given by
\begin{align*}
	X_R(t) =\frac{\beta}{2\pi}\qty(1,\sinh(\frac{2\pi}{\beta}t),\cosh(\frac{2\pi}{\beta}t)) 
\end{align*}
A massless particle coming in at $t = 0$ carries $SL(2)$ charge
\begin{align*}
	q = \frac{\Delta S}{2\pi}\qty(1,\sinh(\frac{2\pi}{\beta}t_m),\cosh(\frac{2\pi}{\beta}t_m))
\end{align*}

Before the particle comes in, the horizon is given by $(1,Y^0,Y^1 = Y^0)$. The particle trajectory is given by $q\cdot Y = 0$. The intersection of the particle trajectory with the horizon is given by $P = (1, e^{\frac{2\pi}{\beta}t_m},e^{\frac{2\pi}{\beta}t_m})$.

After the spin comes in, the new boundary trajectory is given by
\begin{align*}
	\tilde X_R(t) =\ &\frac{\beta}{2\pi}\qty(1-\frac{\Delta S}{S-S_0})\qty(1,\sinh(\frac{2\pi}{\beta}t), \cosh(\frac{2\pi}{\beta}t))\\
	&\ +\frac{\beta}{2\pi}\frac{\Delta S}{S-S_0}\qty(\cosh(\frac{2\pi}{\beta}(t-t_m)),\sinh(\frac{2\pi}{\beta}t_m),\cosh(\frac{2\pi}{\beta}t_m))\\
	&\ +\frac{\beta}{2\pi}\frac{\Delta S}{S-S_0}\frac{2\pi}{\beta}(t-t_m)\qty(0,\cosh(\frac{2\pi}{\beta}t), \sinh(\frac{2\pi}{\beta}t))\\
\end{align*}
\begin{align*}
	-\tilde X_R(t)\cdot P 
	=\ &-\frac{\beta}{2\pi}\qty(1-\frac{\Delta S}{S-S_0})\qty(e^{-\frac{2\pi}{\beta}(t-t_m)}-1)-\frac{\beta}{2\pi}\frac{\Delta S}{S-S_0}\qty(1-\cosh(\frac{2\pi}{\beta}(t-t_m)))\\
	&\ \ \ \ +\frac{\beta}{2\pi}\frac{\Delta S}{S-S_0}\frac{2\pi}{\beta}(t-t_m)e^{-\frac{2\pi}{\beta}(t-t_m)}\\
	\approx\ & \frac{\beta}{2\pi}\qty(1-\frac{\Delta S}{S-S_0})\qty[1+\frac{\beta}{2\pi}\qty(\cosh(\frac{2\pi}{\beta}(t-t_m))-1)]
\end{align*}
The geodesic distance
\begin{align*}
	d = \text{const}+\log(1+\frac{\beta}{2\pi}\qty[\cosh(\frac{2\pi}{\beta}(t-t_m))-1])
\end{align*}

\bibliographystyle{apsrev4-1long}
\bibliography{main}

\end{document}